\documentclass[journal]{IEEEtran}

\ifCLASSINFOpdf

\else

\fi

\usepackage{graphicx}
\usepackage{amsmath}
\usepackage{amssymb}
\usepackage{multirow}

\usepackage{floatrow}
\newfloatcommand{capbtabbox}{table}[][\FBwidth]

\begin{document}
%
\title{DR-GAN: Conditional Generative Adversarial Network for Fine-Grained Lesion Synthesis on Diabetic Retinopathy Images}

\author{Yi~Zhou,~\IEEEmembership{Member,~IEEE},
        Boyang Wang,
        Xiaodong He,
        Shanshan Cui
        and~Ling Shao,~\IEEEmembership{Senior~Member,~IEEE}
\thanks{Corresponding author: \textit{Yi Zhou}.}
\thanks{Y. Zhou, B. Wang, S. Cui and L. Shao are with the Inception Institute of Artificial Intelligence, Abu Dhabi, United Arab Emirates, e-mail: (\{yizhou.szcn, natebywang\}@gmail.com, shanshan.cui@g42.ai, ling.shao@ieee.org)}
\thanks{X. He is with Bodhi Laboratory, Beijing BeYes Technology Co. Ltd. Beijing, China. (email: seariver@foxmail.com)}
}

\markboth{Submitted to IEEE Journal Biomedical and Health Informatics, VOL. XX, NO. XX, XXXX 2020}%
{Zhou \MakeLowercase{\textit{et al.}}: DR-GAN: Conditional Generative Adversarial Network for Fine-Grained Lesion Synthesis on Diabetic Retinopathy Images}
%



\maketitle

\begin{abstract}
Diabetic retinopathy (DR) is a complication of diabetes that severely affects eyes. It can be graded into five levels of severity according to international protocol. However, optimizing a grading model to have strong generalizability requires a large amount of balanced training data, which is difficult to collect, particularly for the high severity levels. Typical data augmentation methods, including random flipping and rotation, cannot generate data with high diversity. In this paper, we propose a diabetic retinopathy generative adversarial network (DR-GAN) to synthesize high-resolution fundus images which can be manipulated with arbitrary grading and lesion information. Thus, large-scale generated data can be used for more meaningful augmentation to train a DR grading and lesion segmentation model. The proposed retina generator is conditioned on the structural and lesion masks, as well as adaptive grading vectors sampled from the latent grading space, which can be adopted to control the synthesized grading severity. Moreover, a multi-scale spatial and channel attention module is devised to improve the generation ability to synthesize small details. Multi-scale discriminators are designed to operate from large to small receptive fields, and joint adversarial losses are adopted to optimize the whole network in an end-to-end manner. With extensive experiments evaluated on the EyePACS dataset connected to Kaggle, as well as the FGADR dataset, we validate the effectiveness of our method, which can both synthesize highly realistic ($1280 \times 1280$) controllable fundus images and contribute to the DR grading task.

\end{abstract}


\IEEEpeerreviewmaketitle

\section{Introduction}
Diabetic retinopathy (DR) is a common disease causing vision loss or even blindness among people with diabetes. Human ophthalmologists usually identify and grade DR severity based on the type and number of related lesions. According to the international protocol \cite{gulshan2016development,drgrading}, DR severity can be graded into five levels: normal, mild, moderate, severe non-proliferative diabetic retinopathy (NPDR) and PDR. The related lesions consist of hard exudates, soft exudates, hemorrhages, microaneurysms, laser marks, proliferate membranes, etc. It is time-consuming and difficult even for ophthalmologists to diagnose DR, so automatic DR grading models \cite{seoud2015automatic,pratt2016convolutional,yang2017lesion,he2019dme} have begun to be explored over the past decades.


Several previous works, \cite{wang2017zoom,lin2018framework,Zhou_2019_CVPR} adopt deep models to implement DR grading and obtain substantial improvement over other methods. Compared with handcrafted feature extraction and traditional machine learning methods, deep convolutional neural networks (CNNs) have achieved great success for many vision tasks, such as image classification \cite{he2016deep}, object detection \cite{liu2018deep}, semantic segmentation \cite{garcia2017review,fan2020inf} and image synthesis \cite{yi2018generative}. Training an effective deep CNN model usually requires a large amount of diverse and balanced data. However, the DR data distribution over different grades is extremely imbalanced since abnormal fundus images only make up a small portion. For example, in the largest public DR dataset, EyePACS \cite{kaggle}, images of DR levels 3 and 4 only account for 2.35\% and 2.16\% of the total, respectively, while normal images of level 0 account for 73.67\%. Adopting such imbalanced data will make the model less sensitive to samples with higher DR severity levels and lead to overfitting. Although common data augmentation methods such as flipping and random cropping and rotation can mitigate the problem, the poor diversity of samples from those levels still limits model performance. Thus, in this paper, we propose an image generation model that synthesizes more miscellaneous DR images with different grading levels, and use these generated images to help train a grading model.

Generative adversarial networks (GANs) \cite{goodfellow2014generative} are widely used in image generation tasks. The GAN framework usually consists of a generative model $G$ and a discriminative model $D$ competing against each other in a min-max game, which has led to great progress in synthesizing photorealistic images. Specifically, one neural network tries to generate realistic data, while the other tries to discriminate between real and synthesized data. DCGAN \cite{radford2015unsupervised} extends GAN by using a deconvolutional layer, which acts as an upscaling operation to transform low-resolution images into higher-resolution ones. CGAN \cite{Mirza2014ConditionalGA} aims to concatenate a one-hot vector with the random noise vector to build conditions into the generator. Moreover, CycleGAN \cite{CycleGAN2017} performs unpaired image-to-image translation from a source domain to a target domain. BigGAN \cite{brock2018large} is an approach that pulls together various of the best recent practices for training class-conditional images and scaling up the batch size and number of model parameters. The result is the routine generation of both high-resolution and high-fidelity images. These well-designed conditional GAN frameworks inspired us to propose a retina generator that can synthesize realistic high-resolution images.

Specifically, our proposed model consists of a high-resolution retina image generator conditioned on vessel, optic disk and lesion masks, a multi-scale spatial and channel attention, multi-scale discriminators with multi-task learning losses, and an adaptive grading manipulation module to control the severity level of DR images synthesized by the generation model. The main contributions of our paper are as follows:

\textbf{1.} A conditional high-resolution image generator is proposed to synthesize retina images with controllable lesion and grading information. In addition to the normal encoder-decoder design, the multi-scale spatial and channel attention and progressive generation design aims to synthesize more realistic local details. Moreover, multi-scale discriminators are optimized by the adversarial loss, feature matching loss and grading classification loss, simultaneously.

\textbf{2.} 	Adopting real DR images with available grading annotations, we learn different latent grading spaces for randomly sampling adaptive grading vectors. These vectors can be regarded as grading embeddings that help manipulate multi-scale synthesis blocks for more effective generation.

\textbf{3.} Both qualitative and quantitive experiments are conducted to evaluate our model. Not only are the fidelity and diversity of the generated samples promising, but we also find that the synthesized images can be used for data augmentation to train better DR grading and lesion segmentation models. Extensive ablation studies and comparison experiments on Kaggle EyePACS \cite{kaggle} and the FGADR \cite{zhou2020benchmark} dataset demonstrate the effectiveness and superiority of our method.

Our preliminary work was presented in \cite{Zhou_2019_DRGAN}. \textbf{Compared to the previous work, in this paper, we make five major extensions.} \textbf{1)} A multi-scale spatial and channel attention (SCA) module is proposed in the synthesis blocks to better enhance the generated quality of fine-details. Then, the grading performance can be improved as well. \textbf{2)} We introduce two more important lesions: laser marks and proliferate membranes, which are particularly related to the identification of level 3 and 4 DR images. We use the weakly-supervised trained activation maps as the lesion mask inputs for these two lesions since we do not have their annotated ground-truths. \textbf{3)} The contours of optic disks in fundus images synthesized by our previous DR-GAN model are usually blurred. In this version, an optic disk mask is added to address this problem and make the generated images more realistic. \textbf{4)} A newly proposed large pixel-level annotated dataset (FGADR \cite{zhou2020benchmark}) is adopted to train the DR-GAN. The performance has been further improved. \textbf{5)} More solid evaluation experiments and results are added, including human evaluation, Frechet inception distance (FID) and Sliced wasserstein distance (SWD) for synthesis fidelity, and true positive rate per class for DR grading.

\section{Related Work}

\textbf{GANs in Medical Image Synthesis:}
Using GANs for medical image synthesis \cite{medicalGAN} could potentially address the shortage of large and diverse annotated databases. Methods have been proposed for a variety of medical imaging domains such as computed tomography (CT) \cite{Nie2016MedicalIS,ruan2020mb,xu2020contrast}, magnetic resonance imaging (MRI) \cite{Jiang2018TumorAwareAD,zhou2020hi,zhao2020tripartite}, and chest X-rays \cite{efficientGAN}. For example, CT imaging causes the risk of cancer due to radiation exposure. A cascade of 3D fully convolutional networks was presented in \cite{Nie2016MedicalIS} to synthesize CT images from MR acquisitions. A pixel-wise reconstruction loss and an image gradient loss were adopted for generation in addition to adversarial learning. Mahapatra et al. \cite{efficientGAN} exploited the conditional GAN and Bayesian neural networks for active learning to synthesize chest X-ray images based on segmented masks. Moreover, positron emission tomography (PET) images are often adopted in diagnosis and staging in oncology but are expensive and radioactive. Thus, a cascade of two conditional GANs was proposed in \cite{MyelinGAN}, where the generators are based on 3D U-Nets and the discriminators are based on basic 3D CNNs, to synthesize PET images from various MR volumes.

\textbf{GANs in Retinal Image Synthesis:}
Recently, some researchers have also exploited GANs to synthesize retinal fundus images. Costa \textit{et al.} \cite{costa2017towards} first adopted a U-Net architecture to transfer vessel segmentation masks to fundus images using a vanilla GAN architecture. However, the generated samples have block defects and do not have controllable grading information. Tub-sGAN \cite{zhao2018synthesizing} was proposed to extend style transfer to the generator and thus increase the diversity of synthesized samples. Though somewhat successful, the DR-related lesions and physiological retina details cannot be synthesized clearly. More recently, Niu \textit{et al.} \cite {niu2018pathological} tried to generate fundus images given pathological descriptors and vessel segmentation masks. The position and quantity of lesions can be adjusted. However, this method only introduced lesion manipulation, without considering a global representation for discriminating grades. The synthesized images still need to be evaluated by ophthalmologists to determine whether they are sufficiently gradable to benefit a grading model. In this work, our generation model can be manipulated with arbitrary grading and lesion information to synthesize corresponding high-resolution images. Thus, the generated samples can be directly exploited to help train DR grading models and improve the accuracy.

\section{Proposed Methods}

\begin{figure*}[h]
\begin{center}
\includegraphics[width=1.0\linewidth]{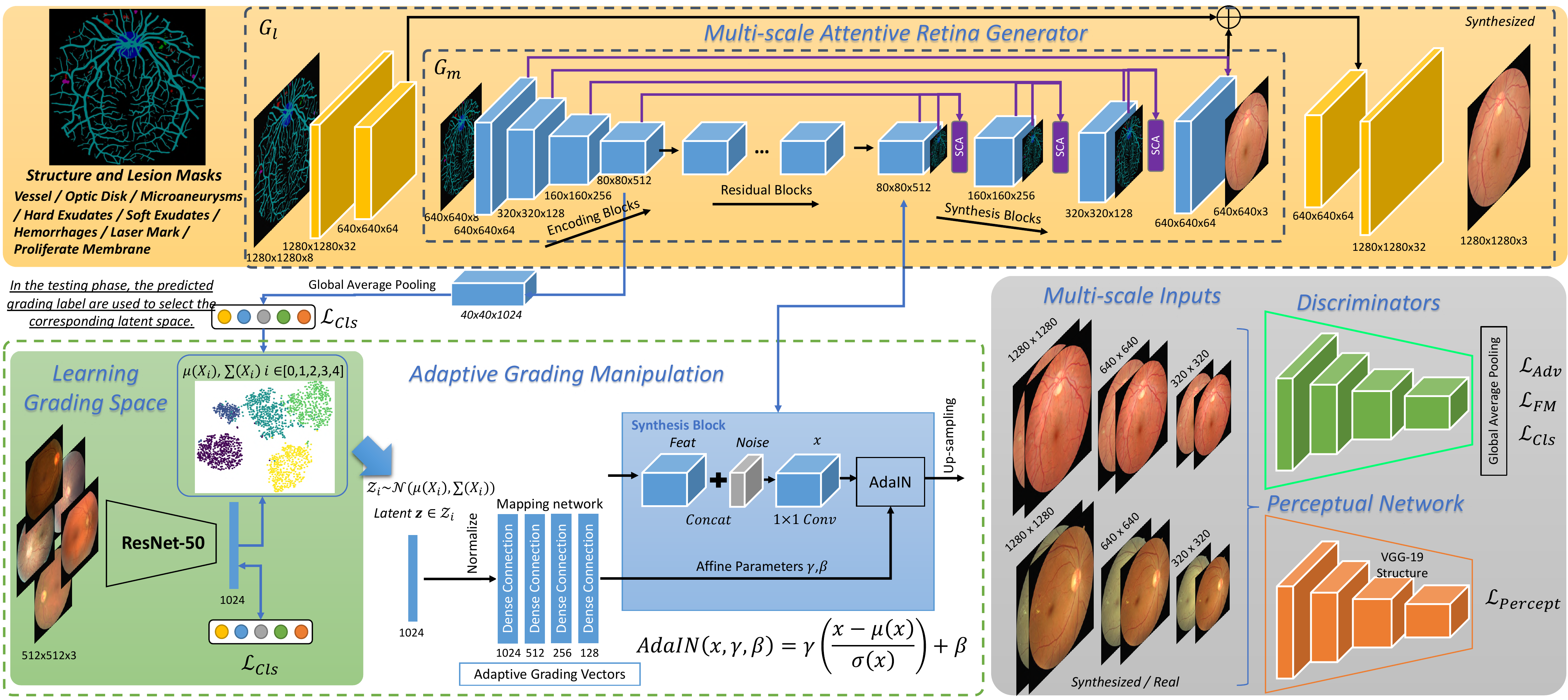}
\end{center}
   \caption{Pipeline of DR-GAN. The trunk of the multi-scale attentive generator in yellow is based on the input of the structural and lesion masks. An adaptive grading manipulation module is designed to be integrated with the synthesis blocks. The gray box shows the multi-scale discriminators for multi-task optimization. The section in green is proposed to learn latent grading spaces for sampling grading embeddings.}
\label{fig:system}
\end{figure*}

\subsection{High-Resolution DR Image Generation}
The resolution requirement for medical images is high since some lesions tend to appear in extremely small regions, such as microaneurysms in fundus images. Inspired by \cite{wang2018high}, we design a two-stage encoder-decoder model to synthesize $1280 \times 1280$ images. As illustrated in Fig.~\ref{fig:system}, the building blocks in blue of the retina generator, denoted as $G_m$, aim to generate a resolution of $640 \times 640$. Then, the blocks in yellow, denoted as $G_l$, can further synthesize more realistic local details to increase the resolution to $1280 \times 1280$. 

$G_m$ consists of three components: encoding blocks, residual blocks and synthesis blocks. The encoding blocks employ a fully-convolutional module with four convolutional ($Conv$) layers. The kernel size of the first $Conv$ layer is set as 7 and that of the remaining $Conv$ layers is 3. We configure the convolutional operation with a stride of 2 instead of using pooling for downsampling. The padding type is set as the same and a ReLU activation and batch normalization are adopted after each layer. The residual blocks increase the depth of the network and are proposed to learn better transformation functions and representations using a deeper perceptron. Finally, the synthesis blocks are the most significant component, whose basic units employ transposed convolutional operations. We embed an adaptive grading manipulation operation into these blocks, which is explained in Sec. 2.3. The hyper-parameter settings of the synthesis blocks are similar to the encoding blocks. In this extension work, we design a multi-scale spatial and channel attention (SCA) module, inspired by \cite{Fu_2019_CVPR}, in the synthesis blocks to better enhance the generation quality of fine-details. This is explained in the next subsection.

$G_l$ has a much simpler design, only including two $Conv$ layers and two corresponding transposed $Conv$ layers. The input to the first transposed $Conv$ layer is the element-wise sum of the feature maps of the last $Conv$ layer of $G_l$ and the feature maps of the last transposed $Conv$ layer of $G_m$. Such a design helps $G_l$ directly inherit the learned global features from $G_m$ and further progressively synthesize local details based on mask inputs, with higher resolution. Please note that $G_m$ is first pre-trained and then $G_l$ is added for fine-tuning the whole generator.

The proposed retina generator for DR images is conditioned on the input of structural and lesion masks. We adopt a U-Net architecture \cite{ronneberger2015u} to train the vessel, optic disk and various lesion segmentation models. Thus, the pairs of real fundus images and corresponding structural and lesion masks can be adopted to train the generator. Since only a small amount of public pixel-wise annotated data is available for training the vessel (DRIVE \cite{staal2004ridge}), optic disk and lesion (IDRID \cite{porwal2018indian}) segmentation models, we use the trained model to predict masks on the large-scale DR dataset (EyePACS \cite{kaggle}) and then adopt them to train our generator. In this extended work, we adopt the FGADR dataset with more pixel-level annotations to improve segmentation for better lesion masks. Although the predicted masks are not the real ground-truths of data from EyePACS, the noise is tolerable and their large-scale amount does contribute to the synthesis performance.

\subsection{Multi-Scale Spatial and Channel Attention Module}
For the retina generator, proposed in our preliminary work \cite{Zhou_2019_DRGAN}, the architecture follows the common encoder-decoder framework. However, it is unable to effectively preserve small details, such as tiny lesions and fine blood vessels, from the input of masks, and synthesize them clearly in the output fundus images. Thus, in this extension work, we propose a spatial and channel attention module in the multi-scale synthesis blocks, as shown in purple in Fig.~\ref{fig:system}. After adaptive grading manipulation in each synthesis block (which will be explained in the next section), the corresponding feature maps from the encoding and synthesis blocks are concatenated, together with the resized input of the structural and lesion masks. The combined features are further passed through one convolutional layer which is taken as the input of the spatial and channel attention module. Then, upsampling is conducted for the next synthesis block.

\begin{figure}[t]
\begin{center}
\includegraphics[width=1.0\linewidth]{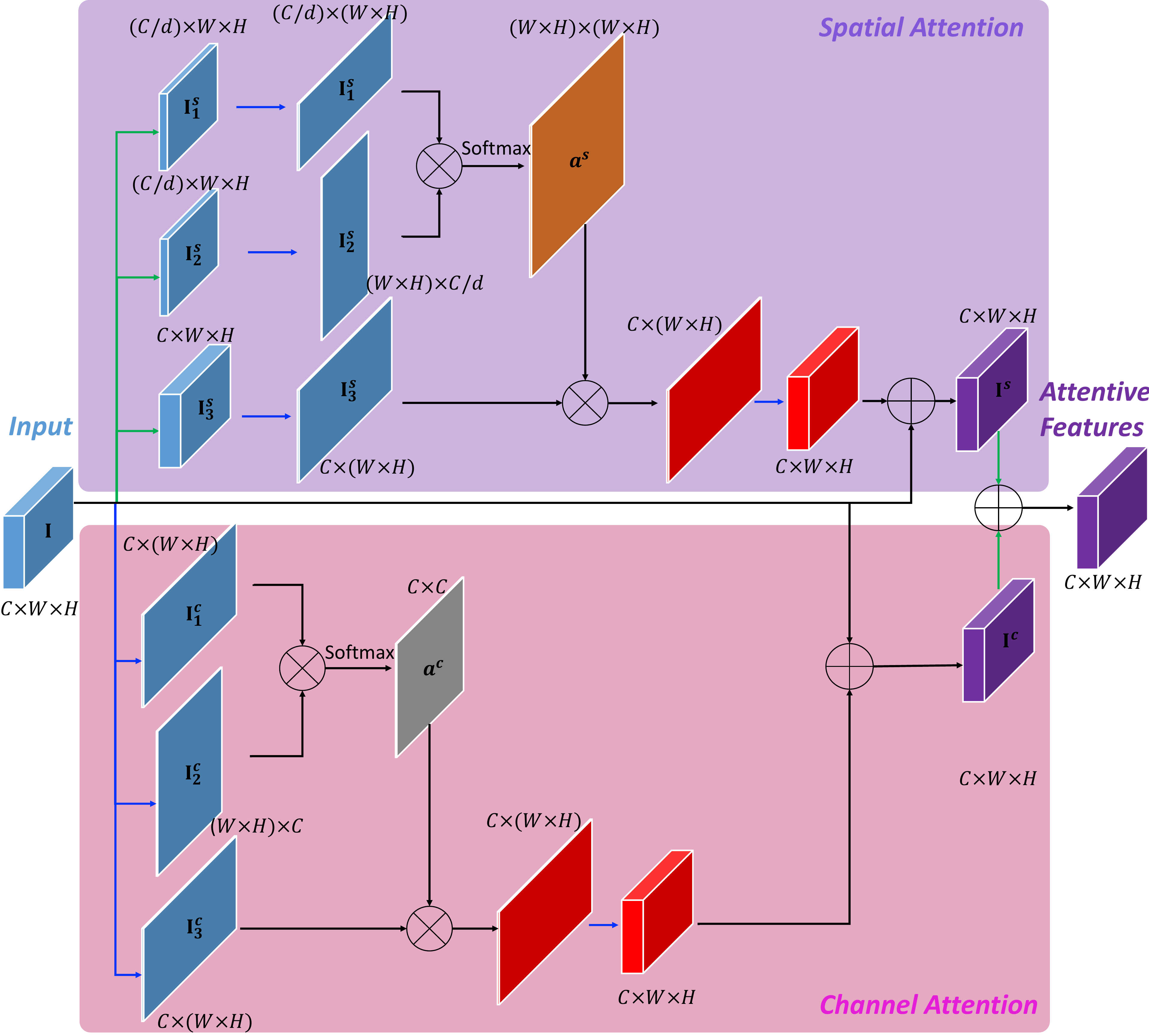}
\end{center}
   \caption{Details of the spatial and channel attention (SCA). The green arrow denotes the convolution operation. The blue one denotes the reshape operation.}
\label{fig:sca}
\vspace{-1ex}
\end{figure}

The network of the spatial and channel attention (SCA) module is illustrated in Fig.~\ref{fig:sca}. It consists of two parts, which operate the spatial and channel attention, respectively. The motivation behind using this self-attention mechanism is to both spatially model richer contextual representations and enhance specific semantics, such as input masks, to improve the dependencies between channel maps. Therefore, small details can be synthesized better.

\textbf{Spatial Attention:} The integrated input feature for the SCA is defined as $\mathbf{I} \in \mathbb{R}^{C \times W \times H}$, where $\{C, W, H\}$ denote the channel, width and height, respectively. Three convolutional layers are operated on the input feature, forming three branches. The first two branches reduce the channel dimension to $C/d$, where $d$ is used for decreasing the computational complexity. Then, $\mathbf{I}^s_1$ and $\mathbf{I}^s_2$ from the two branches are reshaped into $(C/d) \times (W \times H)$ and $(W \times H) \times (C/d)$, separately, and then multiplied. A following Softmax is applied to obtain the spatial attention map $\mathbf{a}^s \in \mathbb{R}^{(W \times H) \times (W \times H)}$. The final spatial attended feature is computed as $\mathbf{I}^s = \mathbf{I} + Reshape(\mathbf{I}^s_3 \mathbf{a}^s))$. In our implementation, we set $d$ to {8, 16, 32} for different scales {$80 \times 80 \times 512$, $160 \times 160 \times 256$, $320 \times 320 \times 128$}, respectively.

\textbf{Channel Attention:} The input $\mathbf{I} \in \mathbb{R}^{C \times W \times H}$ is first reshaped into $\mathbf{I}^c_1 \in \mathbb{R}^{C \times (W \times H)}$ and $\mathbf{I}^c_2 \in \mathbb{R}^{(W \times H) \times C}$, separately. Then, matrix multiplication is conducted with a Softmax to obtain the channel attention map $\mathbf{a}^c \in \mathbb{R}^{C \times C}$. The final channel attended feature is computed as $\mathbf{I}^c = \mathbf{I} + Reshape(\mathbf{a}^c \mathbf{I}^c_3))$. The overall output of the SCA module is the weighted element-wise sum of $\mathbf{I}^s$ and $\mathbf{I}^c$ with one more convolutional layer for each.

\subsection{Multi-Scale Discriminators with Multi-Task Optimization}
To better optimize the high-resolution image generator, the discriminator needs to have receptive fields of different scales to differentiate between real and synthesized images. The most effective method is to design multi-scale discriminators with identical network structures. In this work, we adopt the original size, $1280 \times 1280$, and downsample it to two scales, $640 \times 640$ and $320 \times 320$. The three scales of images are passed forward to the three discriminators $D_n, n \in \{1,2,3\}$. The discriminator applied to the smallest scale provides the largest receptive field to focus on the holistic fundus image structure and some big lesion patterns. In contrast, the one applied to the largest scale provides the smallest receptive field to generate more local details, particularly for small lesion regions. The discriminator structure consists of four convolutional layers with a kernel size of 4 and a stride of 2. A leaky ReLU is used after each layer, with a slope of 0.2. Global average pooling is applied at the end for fitting different scales.

In this work, a multi-task loss is carefully designed for training the generator and the discriminator in an adversarial learning architecture. In addition to employing the standard adversarial loss $\mathcal{L}_{Adv}$, which maximizes the output of the discriminator for generated data, we also adopt the feature matching loss $\mathcal{L}_{Feat\_match}$ \cite{improveGAN} to optimize the generator and match the statistics of features in the intermediate layers of the discriminator. The feature matching loss aims to address the instability of training GANs to prevent the generator from overtraining on the discriminator. Moreover, we also incorporate an auxiliary classification loss $\mathcal{L}_{Cls}$ (adopting focal loss $\mathcal{L}_{focal}$ \cite{lin2017focal} due to the imbalanced data distribution) to enable the discriminator to learn discriminative representations for DR grading, on both the synthesized and real data. For the largest-scale input, an additional perceptual network $F$ based on the VGG-19 backbone is integrated to contribute to the training by $\mathcal{L}_{Perceptual}$.

For the vessel and lesion masks input to the generator $G$, which combines $G_m$ and $G_l$, we fuse them into one conditional map denoted as $\mathbf{c}$. The corresponding real fundus image and grading label are indicated as $\mathbf{x}$ and $\mathbf{y}$, respectively. The overall training loss is defined in the following equation:
\begin{align}
\label{eq:e1}
& \min \limits_{G} ((\sum_{n=1,2,3} \max \limits_{D_n} \mathcal{L}_{Adv}(\mathbf{c},\mathbf{x})) + \lambda_1 \sum_{n=1,2,3} \mathcal{L}_{Feat\_match}(\mathbf{c},\mathbf{x})) \\
\nonumber & + \lambda_2 \min \limits_{G, F} \mathcal{L}_{Perceptual}(\mathbf{c},\mathbf{x}) + \lambda_3 \sum_{n=1,2,3} \min \limits_{D_n} \mathcal{L}_{Cls}(\mathbf{c},\mathbf{x}, \mathbf{y}) \\
\nonumber & = \min \limits_{G} (( \sum_{n=1,2,3} \max \limits_{D_n} (\mathbb{E}[\log D_n(\mathbf{c},\mathbf{x})] + \mathbb{E}[\log(1-D_n(\mathbf{c},G(\mathbf{c})))] \\
\nonumber & + \lambda_1 \sum_{n=1,2,3} (\mathbb{E} [|| D_n^p(\mathbf{c},\mathbf{x}) - D_n^p (\mathbf{c},G(\mathbf{c})) ||_2])) \\
\nonumber & + \lambda_2 \min \limits_{G,F} \mathbb{E} [|| F^q(\mathbf{x}) - F^q (G(\mathbf{c})) ||_1]  \\
\nonumber & + \lambda_3  \sum_{n=1,2,3} \min \limits_{D_n} (\mathcal{L}_{focal}(D_n(\mathbf{x}), \mathbf{y}) + \mathcal{L}_{focal}(D_n(G(\mathbf{c})), \mathbf{y})) \text{,}
\end{align}
where $p$ and $q$ denote the $p^{th}$ and $q^{th}$ intermediate layer in $D_n$ and $F$, respectively. During implementation, we compute all the layers for the feature matching and perceptual losses. Moreover, $\lambda_1$, $\lambda_2$ and $\lambda_3$ control the weights of different losses, and are set as 10, 10 and 1 to make the magnitude of each loss the same size, for the best result. For the discriminator input, a channel-wise concatenation of the conditional maps and the real/synthesized images is conducted.

\subsection{Adaptive Grading Manipulation}
Our final aim is to generate fundus images with controllable DR severity levels which can be used for data augmentation to improve the performance of DR grading models. In this paper, we propose to learn adaptive grading vectors that can be used for manipulation in the synthesis blocks. The overall idea is illustrated in the bottom part of Fig.~\ref{fig:system}. The adaptive grading vectors are learned and sampled from the latent grading spaces.

We first employ ResNet-50 \cite{he2016deep} to train a DR grading model based on the fundus images to extract discriminative features. After visualizing the extracted features using tSNE \cite{maaten2008visualizing}, we can clearly achieve five clusters corresponding to the five DR grading levels. $X_i, i \in [0,1,2,3,4]$ defines the five feature sets, where $X=f_{ResNet-50}(\mathbf{x})$. For each set, we compute the mean $\mu(X_i)$ and variance $\Sigma(X_i)$ to obtain the corresponding normal distribution space. Once the five latent grading distributions are learned, we randomly sample latent vectors $\mathbf{z} \in \mathcal{Z}_i$, where $\mathcal{Z}_i$ is subject to $\mathcal{N}(\mu(X_i),\Sigma(X_i))$. In the training phase, based on the grading ground-truth of an input pair consisting of a mask and real image, the latent vector $\mathbf{z}$ is sampled from the corresponding space.

We take the latent grading vectors as inputs to the synthesis blocks of the generator. Inspired by \cite{karras2018style}, a four-layer non-linear mapping network is first devised to encode the affine transformations, which can benefit the generator manipulated by grading. In each synthesis block, the feature maps $Feat$ are first concatenated with a random noise volume whose number of channels is a quarter that of $Feat$. Then, a $1 \times 1$ $Conv$ is employed to fuse the features. Adaptive instance normalization (AdaIN) \cite{huang2017arbitrary} embeds the grading vectors to transform the original features. The AdaIN is defined as the following function:

\begin{align}
\label{eq:e1}
& \text{AdaIN}(x, \gamma, \beta) = \gamma(\frac{x-\mu(x)}{\sigma(x)}) + \beta \text{,}
\end{align}

where the fused feature $x$ is normalized independently. $\gamma$ and $\beta$ are the scale and bias vectors learned from the adaptive grading vectors, rather than vectors learned from $x$ like in normal batch normalization. In each synthesis block with a different channel-wise dimension of feature maps, the corresponding grading vector is split into $\gamma$ and $\beta$.

\subsection{Testing Phase and Implementation Details}
Based on the input vessel and lesion masks in the testing phase, we need to decide which grading distribution to sample the latent vector from for the grading manipulation. To address this problem, an additional convolution and global average pooling are applied after the last encoding block to further train the grading function. During testing, the predicted grading label is used for the selection. Moreover, once the whole model is trained, we can automatically imitate multiple lesion masks for one vessel mask to synthesize various fundus images with controllable grading labels.

The training scheme for our method consists of two steps.

\subsubsection{Pre-training the Grading Space}
 In the first step, five latent spaces for different grading levels are learned by pre-training a grading model ResNet-50 \cite{he2016deep}. The training and testing for this model exploit the real data from the training and testing set of EyePACS \cite{kaggle}, respectively. The ADAM optimizer is adopted with a learning rate of 0.001 and momentum of 0.5. The mini-batch size is set to 64. We pre-train the model over 30 epochs.

\subsubsection{End-to-End Training}
 Once the pre-training is completed, the proposed DR-GAN is optimized in an end-to-end manner. The number of residual blocks in the generator is set to 7 for the best performance. The ADAM optimizer is adopted with a learning rate of 0.0001 and momentum of 0.5. The mini-batch size is set to 16. The model $G_m$ is first trained over 10 epochs, then $G_l$ is added for fine-tuning over 10 epochs. All the experiments are conducted on an NVIDIA DGX-1. Overall, the training time takes about 85 hours. In the testing phase, our model takes 0.21 seconds to synthesize one image, which is feasible for `online' augmentation.

\section{Experimental Evaluation}
\subsection{Datasets and Pre-processing}
To train the whole DR-GAN model, large-scale data with pixel-wise annotated structural and lesion masks is required. However, the largest public DR dataset is \textbf{EyePACS} \cite{kaggle}, which consists of 35,126 training images and 53,576 testing images only containing grading labels. The images are collected from different sources with various lighting conditions and weak annotation quality. Thus, we adopt a small-scale dataset with pixel-level annotations to train the vessel, optic disk and lesion segmentation models, and then perform inference on EyePACS to obtain masks, which are coarsely used as weak ground-truths for training DR-GAN. In the previous version, we used DRIVE \cite{staal2004ridge} and IDRID \cite{porwal2018indian} (which contain four lesions: microaneurysms, hemorrhages, hard exudates and soft exudates) to train the segmentation models. However, IDRID only has 81 annotated images. In our extended model, we use a newly proposed FGADR dataset with 1,842 images with pixel-level annotations to enhance the training of the segmentation models. Please note that the reason we do not adopt FGADR with real pixel-level annotations to directly train the DR-GAN is because its data scale is small, which will lead to overfitting during training. The EyePACS dataset is large-scale and has higher data diversity. Training using EyePACS can mitigate the overfitting problem and enable DR-GAN to synthesis more diverse retinal patterns.

\begin{figure}[t]
\begin{center}
\includegraphics[width=0.85\linewidth]{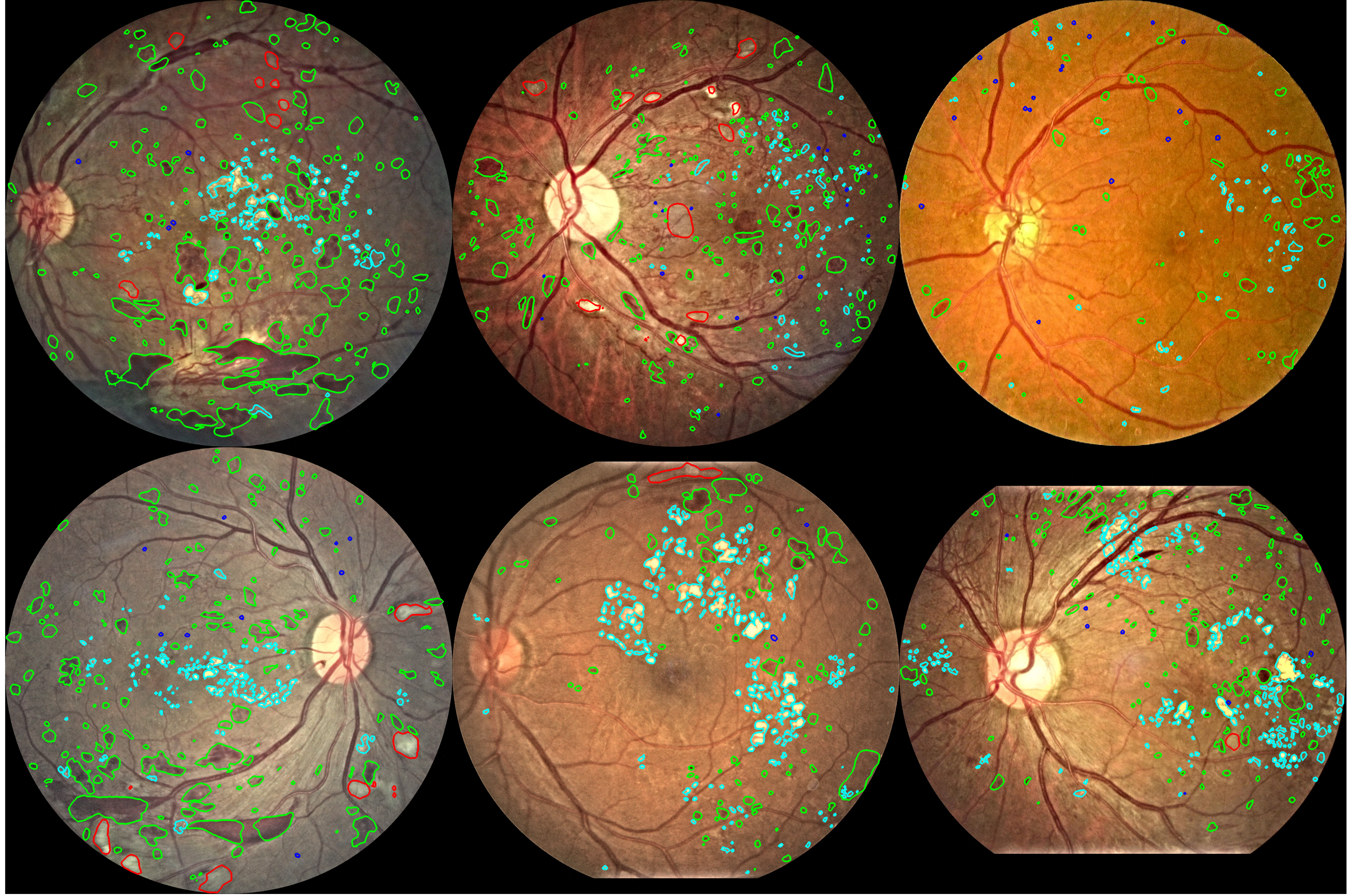}
\end{center}
   \caption{Examples of annotation for the FGADR dataset. The blue, cyan, green and red annotations denote the MA, EX, HE and soft SE, respectively.}
\label{fig:annotation}
\end{figure}

\textbf{FGADR \cite{zhou2020benchmark}:} 
The FGADR dataset consists of a Seg-set containing 1,842 pixel-level annotated images, and a Grade-set containing an additional 1,000 images with only grading labels. The Seg-set has fine-grained pixel-wise lesion annotation as well as DR grading by ophthalmologists. Similar to \cite{porwal2018indian}, microaneurysms (MA), hemorrhages (HE), hard exudates (EX) and soft exudates (SE) are annotated. Moreover, image-level labels for laser marks and proliferate membranes are also provided, since these two lesions are very useful for identifying level 3 and 4 DR but usually appear with holistic features that are difficult to annotate at a pixel level. Fig.~\ref{fig:annotation} provides some examples from the FGADR dataset with annotations. In this extended work, we adopt the FGADR dataset to train the lesion mask segmentation models. We split the 1,842 annotated images into 1,500 images for training and 342 images for testing. The segmentation performance is reported in terms of the area under the curve (AUC) of the precision-recall (PR) curve. We obtain 0.405 for microaneurysms, 0.678 for hemorrhages, 0.798 for hard exudates and 0.692 for soft exudates. Moreover, we also test our model on the IDRID dataset \cite{idrid} for fair comparison and consistently achieve state-of-the-art results: 0.518 for microaneurysms, 0.709 for hemorrhages, 0.892 for hard exudates and 0.758 for soft exudates.

\begin{figure}[t]
\begin{center}
\includegraphics[width=1.0\linewidth]{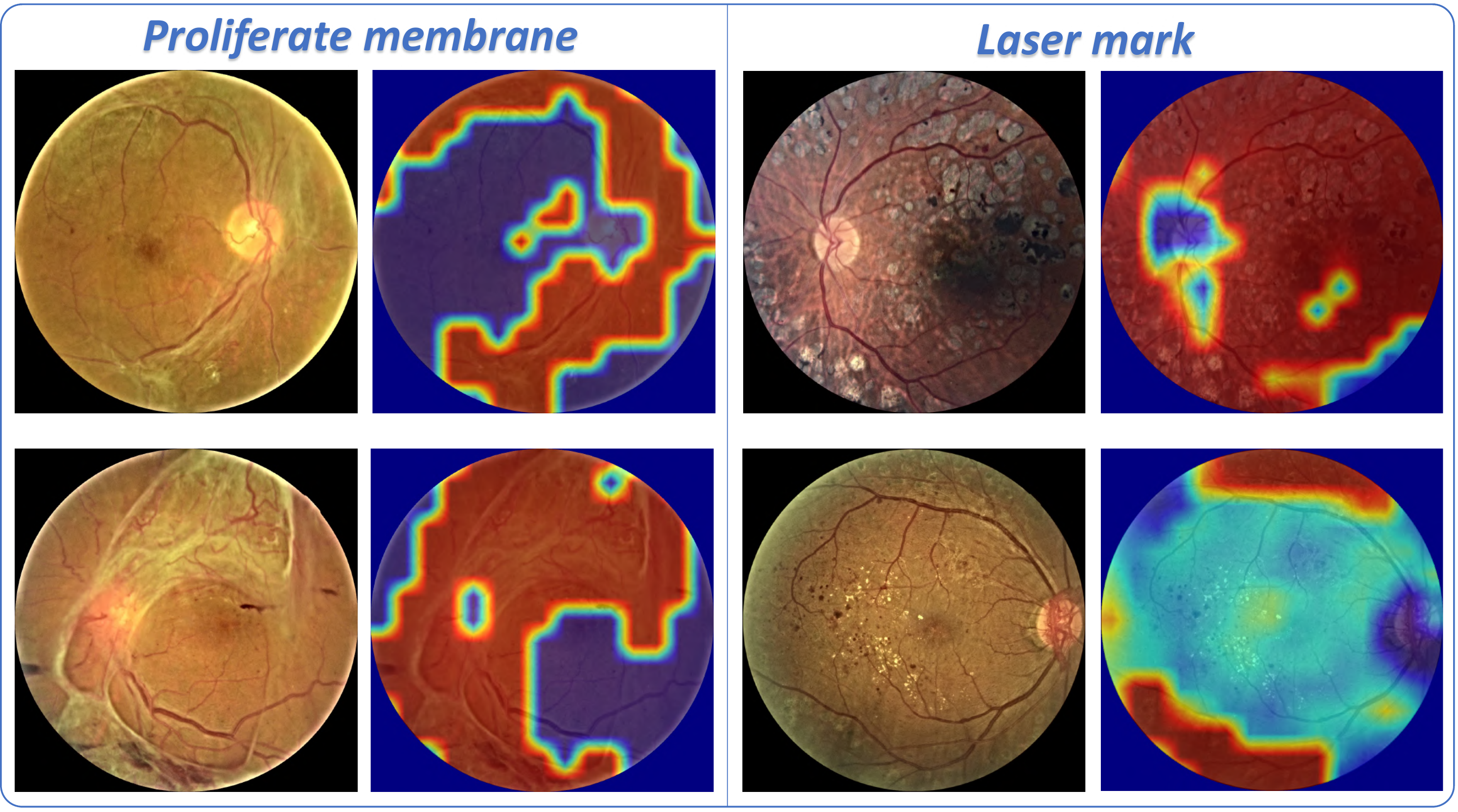}
\end{center}
   \caption{Examples of laser mark and proliferate membrane masks extracted using the weakly supervised method \cite{zhou2016learning}.}
\label{fig:laser}
\end{figure}

\textbf{Laser Mark and Proliferate Membrane Activation Maps:}
Laser marks and proliferate membranes are important lesions that usually appear in severe DR images (i.e. DR-3 and DR-4). However, they are global features, which are not suitable for pixel-wise annotation. Thus, only image-level labels for these two lesions are provided, which indicate whether or not an image has the lesion. When learning discriminative localization \cite{zhou2016learning}, a weakly supervised method is adopted to obtain the activation maps. We select a subset containing all the available images with laser marks and proliferate membranes to train their classifiers. For laser mark classification, 1,450 training images and 398 testing images (60\% of which have positive lesions) are evaluated to get a classification accuracy of 94.97\%. For proliferate membrane classification, 1,382 training images and 398 testing images (20\% of which have positive lesions) are evaluated to get a classification accuracy of 92.21\%. Once the models are trained, the activation maps are extracted for all the images from EyePACS as weak mask ground truths. Fig.~\ref{fig:laser} shows some examples of the weakly supervised activation maps.

\subsection{Qualitative Evaluation of Image Synthesis}
Before evaluating the quantitative improvement of the DR grading performance by data augmentation with synthesized data, we first both qualitatively and quantitively demonstrate the generated image fidelity and evaluate the influence of grading and lesion conditions.

\subsubsection{Performance Demonstration}
\begin{figure*}
\begin{center}
\includegraphics[width=1.0\linewidth]{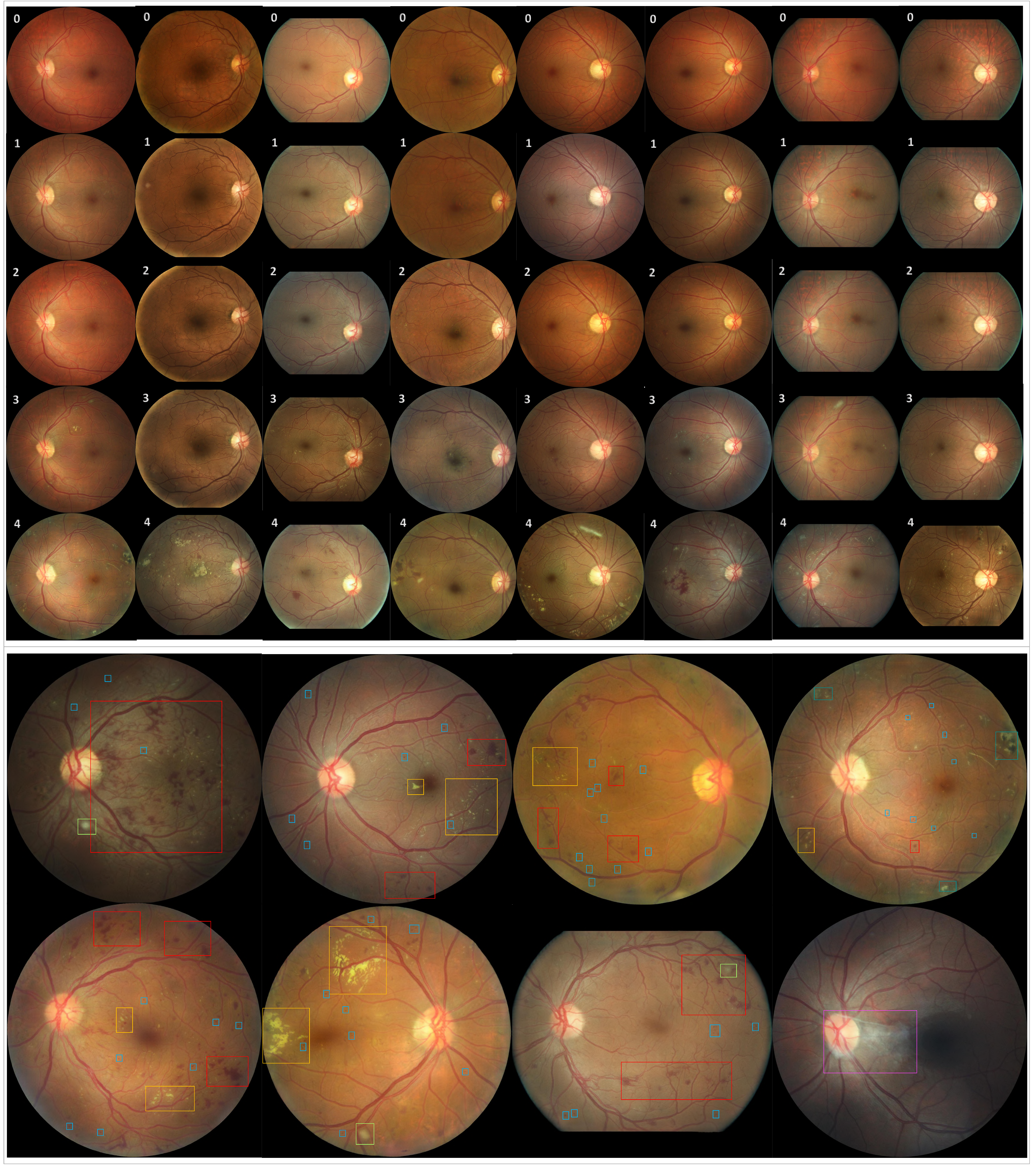}
\end{center}
   \caption{Examples of synthesized images ($\mathbf{1280 \times 1280}$ best viewed zoomed in). Upper part: given an input vessel and optic disk, the corresponding DR images with five different grading levels are generated based on conditional information. Lower part: the detailed synthesized lesion patterns are demonstrated. The red, yellow, light green, blue, dark teal and lavender bounding-boxes indicate hemorrhages, hard exudates, soft exudates, microaneurysms, laser marks and proliferate membranes, respectively.}
\label{fig:synthesis}
\end{figure*}

In this work, we mainly adopt training samples from EyePACS, as well as the extracted vessel, optic disk and lesion masks, to train the model. Once the training is completed, for each structural mask, we can arbitrarily control the quantity and position of the lesion spots within the corresponding masks to synthesize different DR-level images. Specifically, by manipulating the lesion masks, the corresponding grading labels can be coarsely predicted. Thus, we synthesize 10,000 images for each grading level to augment the data. The upper part of Fig.~\ref{fig:synthesis} provides examples of synthesized images with different DR levels, for a given input with vessel and optic disk structures. We find that the fidelity of generated images and the manipulation performance using lesions and grading are highly promising. In addition to the synthesized lesions and structures, we also observe a random color tone variation introduced by AGM. This is because the input latent vector contains not only the discriminative grading features but also other random information such as illumination and color tone. However, this diversity of synthesized data can benefit the  data augmentation performance for DR grading. Finally, the lower part of Fig.~\ref{fig:synthesis} shows more detailed examples of synthesized lesion appearances, including microaneurysms, hemorrhages, hard exudates, soft exudates, laser marks and proliferate membranes.

\begin{figure*}
\begin{center}
\includegraphics[width=1.0\linewidth]{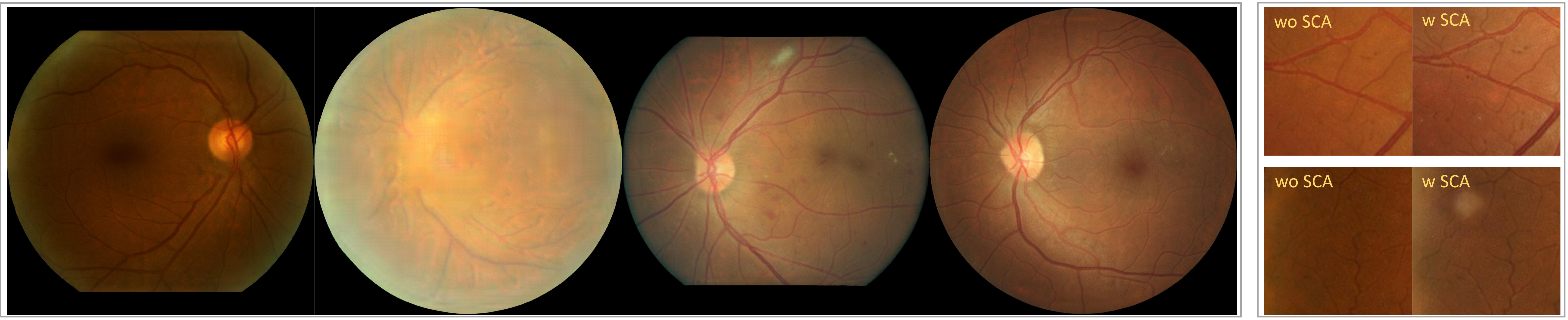}
\end{center}
   \caption{The left four images show failure cases. The first one is a sample with low illumination. The second one is a blurred sample that has failed to synthesize a clear retinal structure. The third one contains a defect in the fovea region. The last one is a grade-1 sample but does not contain any lesions. The right four patches compare the local details synthesized by DR-GAN w/o SCA and DR-GAN w SCA.}
   \vspace{-1ex}
\label{fig:failure}
   \vspace{-1ex}
\end{figure*}

\begin{table*}
\begin{floatrow}
\capbtabbox{
\resizebox{0.55\textwidth}{!}{
\begin{tabular}{c|c|ccc|ccc}
\hline
Methods & Real Images & \multicolumn{3}{c|}{Tub-sGAN \cite{zhao2018synthesizing} Synthesized} & \multicolumn{3}{c}{DR-GAN Synthesized} \\ \hline
Grades  & Fidelity    & Fidelity      & Accuracy 1      & Accuracy 2      & Fidelity     & Accuracy 1     & Accuracy 2    \\ \hline
Grade-0 &   8.95       &    5.82     &    89.3\%    &       -      &      8.33    &    63.2\%    &      -     \\
Grade-1 &    8.85       &   5.69      &      90.8\%       &     -        &     8.04     &      64.5\%    &     -      \\
Grade-2 &   8.91      &      5.33    &       92.2\%      &       -      &     7.98    &     65.8\%     &      -     \\
Grade-3 &    8.87       &     4.96     &       94.9\%      &      -       &      7.75     &       67.7\%     &     -      \\
Grade-4 &   8.86      &     4.99     &       95.1\%      &       -      &     7.73      &       67.9\%     &      -     \\ \hline
Mean    & 8.89        & 5.36          & 92.4\%        &      53.6\%       & 7.97         & 65.8\%       &      85.3\%     \\ \hline
\end{tabular}
 }
}{
 \caption{Human evaluation of the fidelity of synthesized images. \textbf{Fidelity} denotes the realness score, ranging from 1-10. \textbf{Accuracy1} denotes the classification accuracy of real/synthesized samples. \textbf{Accuracy2} denotes the grading accuracy on synthesized samples.}
 \label{tab:human}
}
\capbtabbox{
\resizebox{0.3\textwidth}{!}{
\begin{tabular}{ccc}
\hline
Methods & AVG. FID & AVG. SWD $\times 10^3$ \\ \hline
Tub-sGAN  & 9.38 & 15.62 \\
DR-GAN & 4.53 & 6.43   \\
DR-GAN w SCA &  \textbf{4.24}  &  \textbf{6.17}  \\
w/o Lesion Masks & 8.68 & 16.12 \\
w/o AGM & 6.03 & 8.93  \\
w/o $\mathcal{L}_{Perceptual}$ &  5.76 & 7.74 \\
w/o $\mathcal{L}_{Cls}$ in $D$ & 5.54  & 7.13  \\ \hline
\end{tabular}
}
}{
 \caption{FID and SWD evaluation of the fidelity of the synthesized images. SWD values are multiplied by $10^3$ for better readability.}
 \label{tab:FID}
}
\end{floatrow}
\end{table*}

\subsubsection{Limitations}
Although the DR-GAN model can successfully synthesize high-quality images in most cases, a small number of failure cases still exist due to some limitations. Fig.~\ref{fig:failure} illustrates four main kinds of failure cases observed in all the generated samples. From left to right, the first failure case shows a synthesized sample with low illumination. Our DR-GAN is trained using the EyePACS dataset which contains some samples with poor image quality, such as very low illumination. Since we do not have related labels to manipulate this part of the information in the data distribution, synthesized images will occasionally suffer from this as well. Second, the input of the structural and lesion masks are not real ground truths but inferred from the segmentation models, due to the expense of data annotation. Thus, the performance of the generator will be affected by the quality of the input masks. In the second example, the input vessel and optic disk masks are not clear, leading to a very blurred synthesized image. Third, since the fovea mask is not available during training, sometimes the synthesized fovea region looks unrealistic, as shown in the third case. However, this is a very rare case and does not affect the data augmentation performance. Last, the grading labels of the EyePACS dataset are not absolutely correct, particularly for those DR images with low-level grades (e.g. grade-0 and grade-1). Some DR images are annotated as grade-1 but do not contain any lesions. The last case shows that our model sometimes fails to synthesize corresponding lesions, such as microaneurysms, even though we provide the manipulation information of grade 1. In addition to these four failure cases, our current model cannot manipulate to synthesize intra-retinal microvascular abnormalities (IRMA) or neovascularization (NV), because there are very few real samples containing these two lesions for training. We failed to train good enough segmentation models to provide the masks of these two lesions for the EyePACS dataset.

\subsection{Quantitive Evaluation of Image Synthesis}
To better evaluate the fidelity of the synthesized images, we both conduct a human experts review and compute the Frechet inception distance and sliced wasserstein distance scores.

\subsubsection{Evaluation by Ophthalmologists}
We ask three professional ophthalmologists to independently evaluate the synthesized images. As shown in Table~\ref{tab:human}, three evaluation metrics, including fidelity score, accuracy 1, and accuracy 2, are designed. All the results are obtained by the averaged ratings on the three ophthalmologists. 

In the first experiment, we aim to assess the human-rated image fidelity score of synthesized images. We randomly select 500 images (100 images for each DR level) from the 50,000 generated images, and another 500 real fundus images (100 images for each DR level) for comparison. Human scoring, ranging from 1 to 10, is used to describe the fidelity (i.e. realness) of these images, in terms of details such as vessels, lesion textures and colors, where a higher value is better. We evaluate the realness of real images, since even some real images do not have perfect image quality and the results can show an upper bound performance. Moreover, we evaluate another DR image synthesis method, Tub-sGAN \cite{zhao2018synthesizing}, for comparison. Since Tub-sGAN cannot manipulate the grading information in the synthesis procedure, we use different grading level data to separately train the corresponding models. As shown in Table~\ref{tab:human}, the images synthesized by our method obtain a mean fidelity score of 7.97 (the average result on all five grades), compared to 8.89 for real images, which proves its promising synthesis ability. The images synthesized by the Tub-sGAN receive a poor score of 5.36. Moreover, we observe that the synthesized images with low-level DR grades obtain higher fidelity scores than those with high-level DR grades. This is because synthesizing various lesion appearances of high-level DR images is more challenging and is prone to having defects.

In the second experiment, we mix the 500 synthesized images with the 500 real images and ask ophthalmologists to discriminate whether an image is real or fake. The classification accuracy, denoted as accuracy 1 in Table~\ref{tab:human}, compares the performance by our DR-GAN and Tub-sGAN. The lower the accuracy 1, the better  the quality of the synthesized images. As shown in the results, only 65.8\% of the mixture of real and fake images generated by DR-GAN can be correctly identified by ophthalmologists, compared to 92.4\% for Tub-sGAN. Theoretically, the random guess accuracy should be around 50\%, which means the fidelity of our synthesized images is high. Please note that the rating of real/fake by ophthalmologists in this experiment is independent from the fidelity score in the first experiment.

To better evaluate the grading manipulation ability, we also ask the ophthalmologists to rate the DR grading levels of the 500 images synthesized by DR-GAN and the 500 generated by Tub-sGAN. Accuracy 2 in Table~\ref{tab:human} denotes the grading accuracy on synthesized samples. A grading accuracy of 85.3\% is achieved, which indicates that the images synthesized by DR-GAN contain accurate grading information with corresponding lesion appearances. In contrast, the synthetic image quality of Tub-sGAN is poor. The grading accuracy is only 53.6\%. From the feedback of ophthalmologists, it is even difficult for them to complete the DR rating.

Details of Tub-sGAN \cite{zhao2018synthesizing}: Tub-sGAN adopts a basic generative adversarial framework to synthesize retinal images from vessel masks and random noise. In addition to the generator and discriminator, a VGG-19 network is used for learning style transfer by the style loss and content loss. Since the lesion and grading manipulation is not available in Tub-sGAN, we split all the training data into five parts based on the grading labels to train a model for each grade separately. The testing data is also split for evaluation. Following \cite{zhao2018synthesizing}, images are resized to $512 \times 512$. The dimension of the noise vector is set to 400. Other hyper-parameter training settings also follow \cite{zhao2018synthesizing}.

\subsubsection{Evaluation with Frechet Inception Distance}
To evaluate the performance of image generation models, Frechet inception distance (FID) \cite{heusel2017gans} is usually adopted to measure the similarity between the real image set and the synthesized image set. It has been shown to correlate well with human judgement of visual quality and is most often used to evaluate the quality of samples from generative adversarial networks. FID is calculated by computing the Frechet distance between two Gaussians fitted to feature representations of the Inception network. A lower score indicates better performance. Table ~\ref{tab:FID} shows the average FID scores over the five levels, obtained by different methods, as well as some ablation comparisons, which are explained in sub-section $E.$ Our DR-GAN, without adding the multi-scale SCA module, achieves a score of 4.53, which is a large improvement over Tub-sGAN. With the additional SCA (DR-GAN w SCA), a slight decrease in the FID score can be further obtained. Fig.~\ref{fig:failure} shows this module can improve the realness on small details. Moreover, the result of DR-GAN w/o lesion masks also show that the input of the lesion masks is the most important factor for synthesizing good images. If the input lesion masks are not given, the fidelity of the synthesized lesion patterns is largely decreased.

\subsubsection{Evaluation with Sliced Wasserstein Distance}
In addition to the FID, sliced wasserstein distance (SWD) \cite{Shmelkov_2018_ECCV} is also a suitable metric to evaluate images generated by GANs. SWD is usually adopted to evaluate high-resolution GANs, computing a multi-scale statistical similarity based on local image patches extracted from the Laplacian pyramid representation between generated and real images. Similar to FID, a lower SWD score is better. As shown in Table~\ref{tab:FID}, our DR-GAN with the multi-scale SCA achieves the best score 6.17, which is much lower than that obtained when removing the input lesion masks. Moreover, compared with the DR-GAN without AGM, the SWD score can also be decreased by 2.5, demonstrating that the adaptive instance normalization by the grading vectors can contribute to the synthesis blocks.

\begin{table*}
\begin{floatrow}
\capbtabbox{
\resizebox{0.6\textwidth}{!}{
\begin{tabular}{c|cc|cc|cc|cc|cc}
\hline
\multirow{2}{*}{Data} & \multicolumn{10}{c}{\textbf{Real-1}: training set of EyePACS. \textbf{Real-2}: testing set of EyePACS.} \\
 \multicolumn{1}{c|}{}          & \multicolumn{10}{c}{\textbf{Real-3}: Grade-set of FGADR. \textbf{Fake}: Synthesized 50,000 images (10,000 for each grade).} \\ \hline
Training & \multicolumn{2}{c|}{Real-1}        & \multicolumn{2}{c|}{Real-1} & \multicolumn{2}{c|}{Real-1 + Fake} & \multicolumn{2}{c|}{Real-1} & \multicolumn{2}{c}{Real-1 + Fake}\\
Testing  & \multicolumn{2}{c|}{Fake} & \multicolumn{2}{c|}{Real-2} & \multicolumn{2}{c|}{Real-2}      & \multicolumn{2}{c|}{Real-3} & \multicolumn{2}{c}{Real-3}          \\ \hline
Methods       & Acc.            & Kappa          & Acc.        & Kappa       & Acc.               & Kappa     & Acc.        & Kappa       & Acc.               & Kappa        \\ \hline
VGG-16        & 87.72           & 86.30          & 84.92       & 82.13       & 86.48              & 84.69      &   81.39     &    75.42    &       82.45        &   76.61     \\
ResNet-50     & 89.45           & 88.11          & 86.24       & 83.82       & 88.06              & 85.90      &    82.96    &    76.53    &       84.03        &   77.85    \\
Inception-v3  & 88.37           & 87.20          & 85.80       & 83.44       & 87.63              & 85.44      &    82.03   &    75.92    &     83.07         &    77.02    \\
Zoom-in \cite{wang2017zoom}       & 89.66           & 88.41          & 86.51       & 85.19       & 88.53              & 86.63       &   83.12    &    76.82    &      84.25      &   77.91    \\
AFN \cite{lin2018framework}          & 90.46           & 89.15          & 87.64       & 85.78       & 89.16              & 87.07        &   83.86     &    77.54    &       84.97     &    78.60     \\ \hline
\end{tabular}
 }
}{
 \caption{Evaluation (\%) of augmentation by synthesized data. Please note that FGADR-Grade-set is only used for testing to validate generalization ability.}
 \label{tab:tb1}
}
\capbtabbox{
\resizebox{0.32\textwidth}{!}{
\begin{tabular}{c|c|c}
\hline
Training & Real   & Real+Synthesized \\
Testing  & Real   & Real             \\ \hline
Grade-0  & 0.9376 & 0.9514           \\
Grade-1  & 0.4723 & 0.4901           \\
Grade-2  & 0.7624 & 0.7771           \\
Grade-3  & 0.8411 & 0.8655           \\
Grade-4  & 0.8439 & 0.8622           \\ \hline
\end{tabular}
}
}{
 \caption{TPR per class for evaluating the effectiveness of synthesized data.}
 \label{tab:tb2}
}
\end{floatrow}
\end{table*}

\subsection{Data Augmentation by Synthesis for DR Grading}
Our biggest concern is to evaluate whether the synthesized data can mitigate the unbalanced data distribution over different grading levels and be beneficial for training grading models. We train the baselines for the grading model, which adopt three different classic backbones, VGG-16, ResNet-50 and Inception-v3, with and without using the synthesized data for augmentation. Two state-of-the-art approaches \cite{wang2017zoom} and \cite{lin2018framework} are also compared using the synthesized data.

\begin{figure}[]
\begin{center}
\includegraphics[width=0.75\linewidth]{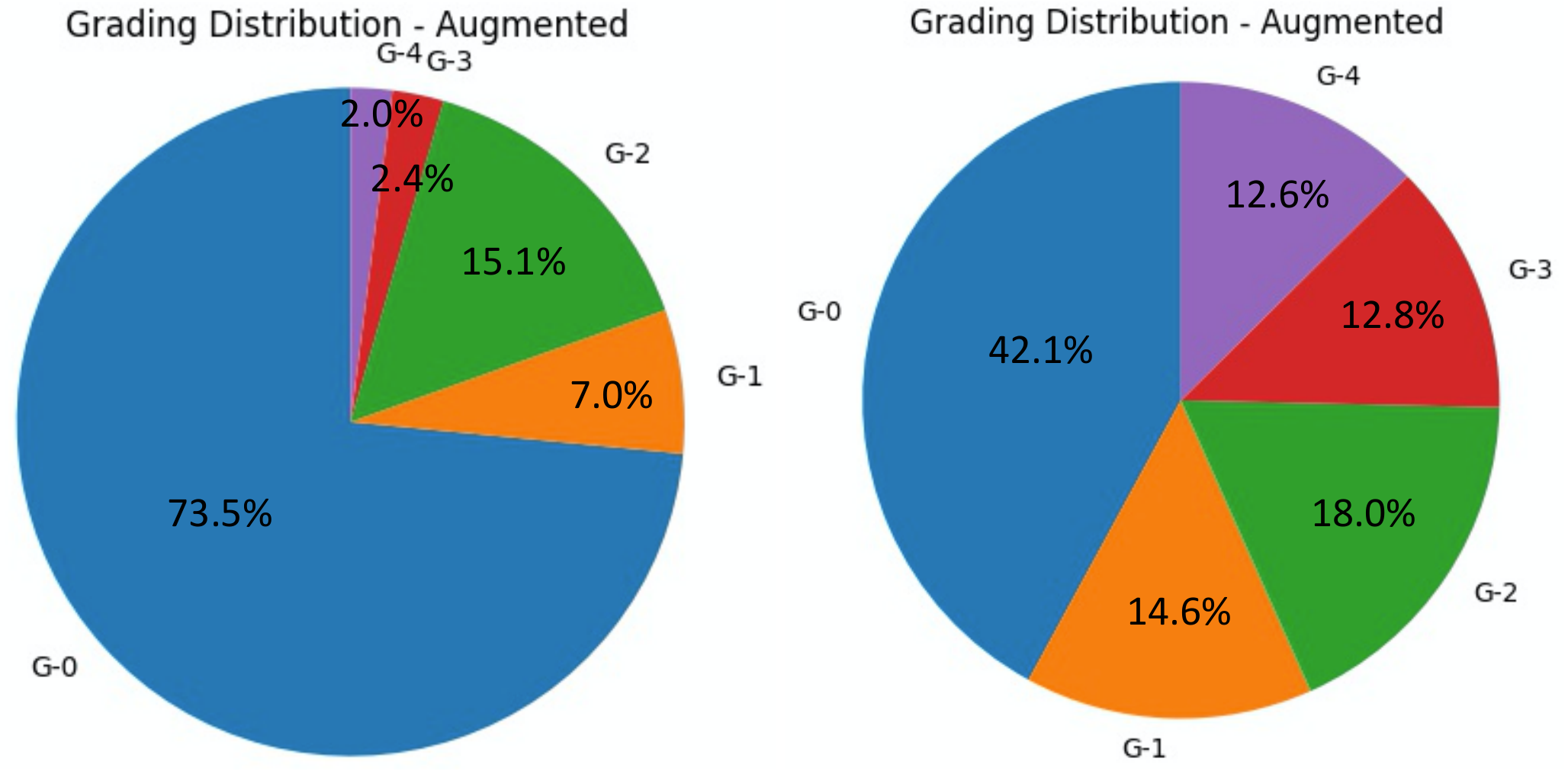}
\end{center}
   \caption{Grading label distributions before and after the data augmentation, for training DR grading models.}
\label{fig:distribution}
\end{figure}

To validate the effectiveness of synthesized data for improving DR grading, we configure three experimental settings. For the first setting, we train the grading models only using the real samples from the EyePACS training set and test them on the synthesized data. As mentioned before, the synthesized data consists of 50,000 images (10,000 images for each grading level). The second setting is a baseline case where the training and testing are conducted on the real training set and testing set of the EyePACS dataset, respectively. For the third one, we combine the training set of EyePACS and the synthesized data for training and evaluate on the real testing set of EyePACS. Fig.~\ref{fig:distribution} compares the grading label distributions of the training data before and after the data augmentation. As illustrated in the Table~\ref{tab:tb1}, both the classification accuracy and the quadratic weighted kappa metric \cite{kaggle} are employed for evaluation. First, adopting the model trained on the real data with grading ground truths, we observe that a highly promising grading performance on the synthesized data can be achieved. A classification accuracy of 90.46\% and kappa value of 89.15\% are obtained by the best-performing AFN \cite{lin2018framework} model. Moreover, for the evaluation on the setting of real test images from the EyePACS, the synthesized data are added into the training set for augmentation. The results show that consistent improvement is achieved over all five compared approaches. The accuracy is increased on average by 1.75\% and the kappa is increased by 1.87\%. To explore whether the data augmentation can generalize well on other datasets, we also test all the models on the Grade-set of FGADR. Please note that we do not perform any fine-tuning using the FGADR data. As shown in Table~\ref{tab:tb1}, the data augmentation by synthesized data increases the accuracy on average by 1.08\% and the kappa by 1.15\%. We believe that once we obtain more training data with more accurate lesion masks, the proposed DR-GAN will be further enhanced and contribute to the grading performance even more significantly.

\begin{figure}[]
\begin{center}
\includegraphics[width=0.9\linewidth]{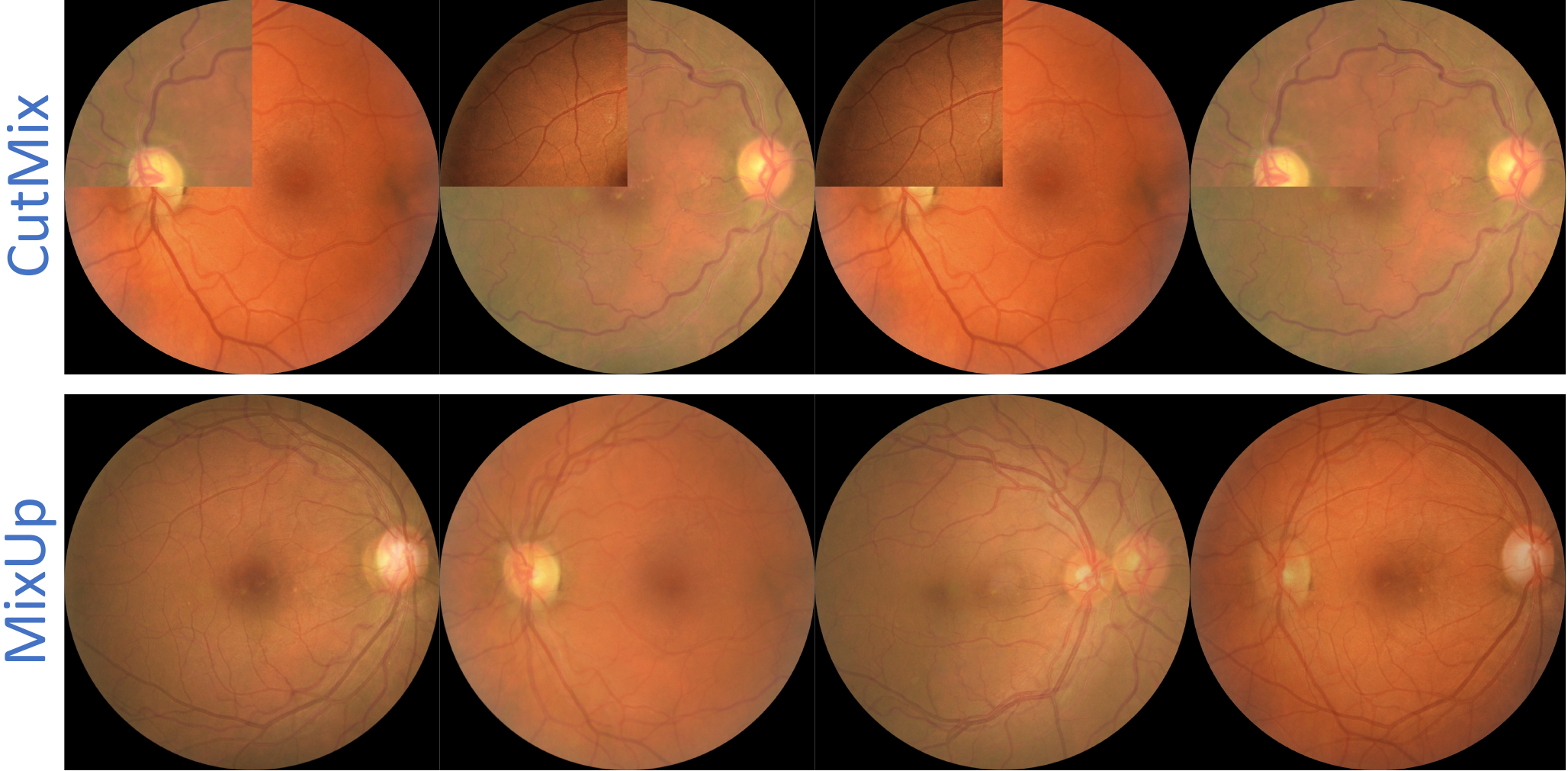}
\end{center}
   \caption{Examples of the generated data by MixUp \cite{zhang2018mixup} and CutMix \cite{yun2019cutmix}.}
\label{fig:aug}
\end{figure}

To address another concern, whether some light-weight augmentation methods can be useful, we adopt MixUp \cite{zhang2018mixup} and CutMix \cite{yun2019cutmix} for data augmentation. For MixUp, two images of the same grade level are mixed by linear interpolation. For CutMix, we cut and paste patches between images of the same grade level. Examples by the two methods are shown in Fig.~\ref{fig:aug}. For both methods, 50,000 images are generated to evaluate the effects of augmentation for DR grading on the EyePACS dataset. The evaluated model is based on the ResNet-50 backbone. However, the average grading accuracy using MixUp decreases by 0.45\% compared to that without augmentation. Only a slight increase of 0.21\% using CutMix is achieved. This is because these two methods cannot synthesize realistic images and also unable to generate the correct number of lesions for the corresponding grade levels.

Except the kappa and average accuracy, we also calculate the true positive rate (TPR) per class to be able to determine whether the synthesized samples improve the performance on each class. We mainly evaluate the effectiveness of adding the fake data into the training set. The fake data are synthesized by DR-GAN with the multi-scale SCA module. The test results on EyePACS are reported in Table~\ref{tab:tb2}. The evaluated model is based on the ResNet-50 backbone. We find that improvements can be achieved over all the classes, particularly for grade-3 and grade-4, due to the lack of real data.

\subsection{Ablation Studies}

To separately evaluate the effectiveness of manipulation by lesion and grading information and the contribution of the multi-loss training, we conduct five ablation studies based on the ResNet-50 baseline. \textbf{Without (w/o) Lesion Masks:} We first study the effect of dropping the input lesion masks and only using the grading manipulation module, by arbitrarily selecting the latent grading space. \textbf{w/o Adaptive Grading Manipulation (AGM):} Oppositely, we investigate the effectiveness of the AGM module by detaching it while keeping the input lesion masks. \textbf{w/o $\mathcal{L}_{Perceptual}$} and \textbf{w/o $\mathcal{L}_{Cls}$} are also explored for their respective contributions to the overall loss functions. \textbf{DR-GAN w SCA:} In this extension work, we add the multi-scale spatial and channel attention module into the generator. We also explore the effectiveness of this design. Table~\ref{tab:ablation} compares the grading performance by each baseline. We find that dropping the input lesion masks significantly affects the grading performance due to the poor quality of synthesized lesion patterns. Thus, augmentation by the generated data cannot contribute to the grading model. Besides, compared with the model without AGM, this design can increase the grading result by a margin of 1.68\% for kappa. The AGM can improve the fidelity and diversity of the synthesized lesion appearances within the corresponding grading levels. Moreover, dropping $\mathcal{L}_{Perceptual}$ or $\mathcal{L}_{Cls}$ in the multi-loss training will both reduce the grading performance. Particularly, for the discriminator without the embedded classification loss, the grading accuracy decreases by 1.17\%, while the kappa value decreases 1.83\%. Finally, with the enhancement by the multi-scale SCA module, not only can the fidelity of synthesized images be improved, but we also obtain a slight increase in the performance of the grading model.

\begin{table}[t]
\caption{Evaluation (\%) of the effectiveness of each module.}
 \label{tab:ablation}
\resizebox{0.85\textwidth}{!}{
\begin{tabular}{c|cc|cc}
\hline
Training & \multicolumn{2}{c|}{Real}        & \multicolumn{2}{c}{Real+Synthesized}  \\
Testing  & \multicolumn{2}{c|}{Synthesized} & \multicolumn{2}{c}{Real}        \\ \hline
Baseline-Res50 & Acc. & Kappa & Acc. & Kappa \\ \hline
DR-GAN &  89.32  &   87.99  &   87.98  &  85.81   \\
DR-GAN w SCA &   89.45  &  88.11  &  88.06   &   85.90   \\  
w/o Lesion Masks &  76.31   &  72.40   &  84.21   &  80.61   \\
w/o AGM &  86.42   &  84.32   &  86.53  &  84.13   \\
w/o $\mathcal{L}_{Perceptual}$ &  87.30   &  86.44   &  86.98   &  84.76  \\
w/o $\mathcal{L}_{Cls}$ in $D$ &  87.54  &   86.09  &  86.81  &  83.98  \\ \hline
\hline
\multicolumn{5}{c}{Effects of the pre-trained grading space for AGM}                             \\ \hline
\multicolumn{1}{c|}{Pre-trained Accuracy} & Acc. & \multicolumn{1}{c|}{Kappa} & Acc. & Kappa \\ \hline
\multicolumn{1}{c|}{75.40}                     &   83.04   & \multicolumn{1}{c|}{81.96}      &   82.95   &   81.48    \\
\multicolumn{1}{c|}{79.26}                     &   86.47   & \multicolumn{1}{c|}{84.31}      &   86.51   &   84.09    \\
\multicolumn{1}{c|}{83.53}                     &   87.91   & \multicolumn{1}{c|}{86.80}      &   87.63   &  85.24     \\
\multicolumn{1}{c|}{86.24}                     &   89.45   & \multicolumn{1}{c|}{88.11}      &   88.06   &   85.90    \\ \hline
\end{tabular}
}
\end{table}

The latent grading space learned from the pre-trained grading model, ResNet-50, plays an important role in manipulating the grading information. To explore the effect of the pre-trained grading space for AGM, we test the pre-trained models with different accuracies to find the critical value needed to obtain a usable latent distribution. Once the training of the ResNet-50 is complete, we select four checkpoints obtaining different grading accuracies: 75.40\%, 79.26\%, 83.53\%, and 86.24\%. The bottom part of Table~\ref{tab:ablation} shows the results of these different pre-trained accuracies. We observe that the performance of AGM is similar to the model \textbf{w/o AGM} when the pre-trained grading accuracy is only 79.26\%. If the pre-trained grading space is worse, with an accuracy of only 75.40\%, the contribution of AGM in systhesizing images is negative. This means that a critical grading accuracy of around 80\% for the pre-trained model is required to generate an effective latent grading space.

\section{Conclusion}
In this paper, we proposed an effective high-resolution DR image generation model which is conditioned on the grading and lesion information. The synthesized data can be used for data augmentation, particularly for those abnormal images with severe DR levels, to improve the performance of grading models. In our future work, more real annotated pixel-level lesion masks will be added for training DR-GAN better.




\ifCLASSOPTIONcaptionsoff
  \newpage
\fi

\bibliographystyle{IEEEtran}
\bibliography{tmibibliography}

\end{document}